\documentclass[a4paper,11pt,twoside]{article}

\usepackage{amsmath, amssymb, amsthm, ascmac}
\usepackage{actuarialsymbol}
\usepackage{color}
\newcommand{\red}{\color[rgb]{0,0,0}}

\usepackage{graphicx}

\usepackage{fourier}

\newtheorem{thm}{Theorem}[section]
\newtheorem{remark}[thm]{Remark}

\numberwithin{equation}{section}
\allowdisplaybreaks

\def\R{\mathbb{R}}

\def\N{\mathbb{N}}

\def\e{{\epsilon}}

\def\d{\delta}
\def\l{\left}
\def\r{\right}
\def\a{\alpha}
\def\b{\beta}

\def\g{\gamma}
\def\th{\vartheta}
\def\s{\sigma}

\def\k{\kappa}
\def\la{\lambda}

\def\F{\mathcal{F}}

\def\P{{\mathbb{P}}}

\def\E{\mathbb{E}}

\def\mb#1{\mbox{\boldmath $#1$}}

\def\I{\mathbf{1}}

\def\wh#1{\widehat{#1}} 
\def\ol#1{\overline{#1}} 
\def\wt#1{\widetilde{#1}} 

\def\Cov{\mathrm{Cov}}
\def\df{\mathrm{d}}

\newcommand{\bi}{\begin{itemize}}
\newcommand{\ei}{\end{itemize}}

\title{Mortality Prediction using Survival Energy Models with Functional Data Analysis}
\author{Daiki Mitsuda\footnote{{\tt numbercity@ruri.waseda.jp }} \\
{\it Graduate School of Fundamental Science and Engineering, Waseda University}\\
Yasutaka Shimizu\footnote{{\tt shimizu@waseda.jp}} \\ 
{\it Department of Applied Mathematics, Waseda University} 
}
\date{\today}

\begin{document}

\maketitle 

\begin{abstract} 
The Survival Energy Model (SEM), as originally introduced by Shimizu {\it et al.} \cite{s20}, is designed to characterize human bioenergetics by employing diffusion processes or inverse Gaussian processes. While parametric models have been employed to articulate the SEM, they exhibit inherent sensitivity in their parameters and hyperparameters, which in turn introduces issues of instability in estimation and prediction. In this paper, we demonstrate that the utilization of functional data analysis techniques for nonparametric estimation and prediction of critical functions within the SEM leads to a substantial enhancement in prediction performance.

\begin{flushleft}
\vspace{1mm}
\end{flushleft}
\end{abstract}

\section{Introduction}\label{sec:intro}
{\it Survival Energy Model (SEM)} is a stochastic model proposed by Shimizu {\it et al.} \cite{s20} for cohort-wise prediction of mortality. 
It assumes that human death is caused by the disappearance of hypothetical `survival energy' and expresses the dynamics as a stochastic process. 

On a stochastic basis $(\Omega, \F, \P; \mathbb{F})$ with the right-continuous filtration $\mathbb{F}:=(\F_t)_{t\ge 0}$, a survival energy process of a certain cohort, say $c$, is defined as a c\`adl\`ag process $X^c=(X^c_t)_{t\ge 0}$. Then the time of death, say $\tau$, is defined as the first-hitting time of $X$ to zero: 
\[
\tau_c:=\inf\{t > 0\,|\, X^c_t < 0\}, 
\]
which is a $\F_t$-stopping time. Thus, the probability of death up to time $t>0$ of the corresponding cohort is given by 
\begin{align}
q_c(t):=\P(\tau_c \le t), \label{mortality}
\end{align}
which is called the {\it mortality function} at cohort $c$. 

The first SEM by Shimizu {\it et al.} \cite{s20} is an {\it inhomogeneous diffusion process}, called {\it ID-SEM}:  
\begin{align}
X^c_t = x_c + \int_0^t U_c(s)\,\df s + \int_0^t V_c(s)\,\df W_s,  \label{ID}
\end{align}
where $x_c >0$ is the `initial energy', $U_c,\,V_c:[0,\infty) \to \R$ are deterministic functions, and $W$ is a Wiener process. 
The model features intuitive modeling by interpreting $U_c(t)\df t$ as representing the average drift at time $t $, and $V_c^2(t)\df t$ as representing its dispersion, which can cause sudden death if it values larger. Note that this SEM has a continuous path, and $X^c_{\tau_c} =0$ almost surely. 

On the other hand, the second SEM by Shimizu {\it et al.} \cite{s23} is 
based on an {\it inverse-Gaussian process},  say $Y^c=(Y^c_t)_{t\ge 0}$: 
\begin{align}
X^c_t = x_c - Y^c_t, \label{IG}
\end{align}
where the probability density of $Y^c_t$ is given by
\begin{align*}
f_{Y_t^c}(y) = {\red \sqrt{\frac{\s_c \Lambda_c^2(t)}{2\pi y^3}} \exp\l( - \frac{\s_c (y - \Lambda_c(t))^2}{2 y}\r), }
\end{align*}
where $\s_c>0$ is a constant and $\Lambda_c$ is a nonnegative increasing function with $\Lambda_c(0)=0$. This SEM is called {\it IG-SEM}, 
and the path may have jumps such that $X^c_{\tau_c} < 0$. In particular, if $\Lambda_c(t) =t$, then $X^c$ is a spectrally negative L\'evy process with infinite activity jumps. Although the interpretation of $\Lambda_c$ is not as intuitively easy as in the case of ID-SEM, $\Lambda_c(t)$ approximately represents the degree of energy change at time $t$, and the larger $\Lambda_c$, the higher the mortality rate in the model.

The utilization of the SEM offers numerous computational advantages in actuarial science, particularly in the context of cohort-specific premium calculations, sensitivity analyses for premiums with respect to the change of mortality, and longevity assessments. This is well-documented in the works of Shimizu {\it et al.} \cite{s23-2} and Shirai and Shimizu \cite{ss22}. 

In employing these models, a suitable choice of $U_c, V_c$, and $\Lambda_c$ is important. Occasionally, the predictive performance can be influenced by the hyperparameters embedded within these functions. While a well-constructed parametric model has demonstrated the potential to outperform not only the classical Lee-Carter model \cite{lc92} but also other established cohort models such as the CBD model by Cairns {\it et al.} \cite{cetal06} and the RH model by Renshaw and Haberman \cite{rh03, rh06} — as detailed in Shimizu {\it et al.} \cite{s20, s23}  numerical comparisons — the converse outcome can easily transpire if executed improperly. The selection of such parametric models requires a universal measure, like an information criterion. However, constructing such a criterion poses challenges since $X^c$ represents a fictitious quantity that was originally unobservable.
To overcome this difficulty, embracing the contemporary paradigm of {\it Functional Data Analysis (FDA)} holds great promise.

In this manuscript, we aim to estimate these functions and generate forecasts for prospective cohorts through the application of the FDA. 
First, we employ nonparametric estimation of the functions $U_c, V_c$, and $\Lambda_c$, using the least-squares method at specific time points $t= t_1,\dots,t_d$. Plotting these estimates with respect to $t$, achieved through appropriate smoothing techniques, furnishes us with the functional data for $U_c, V_c$ and $\Lambda_c$. This process is undertaken for each cohort, facilitating our comprehension of how these functional data evolve with cohort progression. Subsequently, by applying the FDA's guidelines to these developments, we can predict the functions $U_c, V_c$, and $\Lambda_c$ for forthcoming cohorts. 
These predictions are then complemented by our original refinements, culminating in the computation of a predictive mortality function within each SEM.

Our research shows that using SEM predictions with the FDA is better than using parametric frameworks. 
Importantly, the FDA technique we use is a widely accepted and easy-to-use method available in the R programming package.

In Section \ref{sec:fda}, we explain the specific steps for estimating and predicting using the FDA for some critical functions, say `key' functions, that represent each SEM. It is important to note that the predictions from the `key' function determine the projected mortality function. We also provide an overview of {\it Functional Principal Component Analysis (FPCA)} in Subsection \ref{sec:fpca}, as it constitutes a primary tool in our forecasting procedure. 
Section \ref{sec:data} is dedicated to the analysis of real data, sourced from the Human Mortality Database (HMD) \cite{hmd}, which is a prominent open-source repository. Our analysis conclusively demonstrates the superior performance of our proposed methodology in contrast to parametric alternatives.
The paper is closed with concluding remarks, presented in Section \ref{sec:remark}, in which we synthesize our findings and offer insights into the broader implications of our research.

\section{Prediction procedure with FDA}\label{sec:fda}

\subsection{Functional datatization of `key' functions in SEM}\label{sec:key}

Our initial undertaking involves the process of "datatization" of the `key' functions, which constitute unobservable entities within the SEM. The fundamental concept is to estimate these functions at discrete time points for each cohort and subsequently apply a basis expansion to transform them into (continuous) functional data.

Recall the explicit expressions for mortality functions given in \eqref{mortality} in ID- and IG-SEM.  \vspace{1mm}

\noindent {\bf ID-SEM:} The mortality function of \eqref{ID} for cohort $c$ is given as follows (Shimizu {\it et al.} \cite{s20}). 
\begin{align}
q_c^{ID}(t; M_c) = 1 -\Phi\l(\frac{x_c + M_c(t)}{\sqrt{2S_c(t)}}\r) + e^{-\k_c x_c}\Phi\l( \frac{-x_c + M_c(t)}{\sqrt{2S_c(t)}} \r), \label{ID-q}
\end{align}
where $\Phi$ is the cumulative distribution function of the standard normal distribution, 
$M_c(t) = \int_0^t U_c(s)\,\df s$ and $S_c(t) = \frac{1}{2}\int_0^t V_c^2(s)\,\df s$ under the assumption that 
\begin{align}
\frac{M_c(t)}{S_c(t)}\equiv \k_c < 0\ (\mbox{constant}). \label{kappa}
\end{align}
Due to the last assumption, it suffices to know the negative constant $\k_c$ and, e.g., the function $M_c$ (or $S_c$) to have $q_c^{ID}$. 
Therefore, we consider $M_c$ a `key' function in ID-SEM.  \vspace{1mm}

\noindent {\bf IG-SEM:} The mortality function of \eqref{IG} for cohort $c$ is given as follows (Shimizu {\it et al.} \cite{s23}). 
\begin{align}
q_c^{IG}(t;\Lambda_c) = \Phi\l(\sqrt{\frac{\s_c}{x_c}}(\Lambda_c(t) - x_c)\r) - e^{2\s_c \Lambda_c(t)}\Phi\l( - \sqrt{\frac{\s_c}{x_c}}(\Lambda_c(t) + x_c)\r). \label{IG-q} 
\end{align}
In this model, we need to know the constant $\s_c$ and the function $\Lambda_c$, which is the `key' function in IG-SEM. 

\begin{remark}
In this paper, we adopt a fixed approach with respect to the parameters $\kappa_c$ in the ID-SEM and $\s_c$ in the IG-SEM, setting them to specific values that are determined through empirical selection. These values are referenced in \eqref{hyper} within the data analysis in Section \ref{sec:data}. The empirical selection of these values is informed by the outcomes of parametric inference, with the primary objective being the attainment of stability in the estimation of $M_c$ or $\Lambda_c$.
While the determination of $\k_c$ and $\s_c$ remains an important consideration, we do not discuss it further in this paper, but it is worth noting that even if these values are chosen in an ad hoc manner, the predictive efficacy of the mortality function remains good, provided that the `key' functions can be accurately estimated.
\end{remark}

Our primary objective is to acquire the "functional data" pertaining to the `key' functions. We denote an empirical mortality function by $q_c^{DATA}(t)$ $(t=S, S+1, \dots, w;\ c=c_1,c_2,\dots,c_m)$. These empirical functions are derived from data in the Human Mortality Database, as detailed in Shimizu et al. \cite{s20}. Here, $w$ denotes the final age recorded in the Human Mortality Database, and the value of $S$ is determined on a case-by-case basis by the observer.

In the sequel, we will explain the procedure to obtain functional data for $M_c$ in the case of ID-SEM. 
The one for $\Lambda_c$ in the case of IG-SEM is the same. 
Moreover, in practice, we use the conditional version of the mortality function following the manner of Shimizu {\it et al.} \cite{s20}: e.g., in the case of ID-SEM, 
\begin{align*}
q^{ID}_c(t,{\red M_c}|S)&:= \frac{q_c^{ID}(t;{\red M_c}) - q_c^{ID}(S;{\red M_c})}{1 - q_c^{ID}(S;{\red M_c})} = \P(\tau_c \le t|\tau_c > S),\quad t>S,  
\end{align*}
and its empirical version 
\begin{align*}
 q^{DATA}_c(t|S)&:= \frac{q_c^{DATA}(t) - q_c^{DATA}(S)}{1 - q_c^{DATA}(S)}, \quad t>S. 
\end{align*}

\begin{remark}\label{rem:data}
We assume we have mortality data for individuals born in cohort $c_m$ until they reach the age of $w$ years. In other words, we have access to mortality data extending up to the calendar year $c_{m+w}$. Consequently, for cohort $c_{m+L}$, we have mortality data available up to the age of $w-L$ years, and our objective for this specific cohort is the prediction of the mortality function $q_c(t|S)$ for $t> (w-L)\vee S$
\end{remark}

{\bf Smoothing for discretely estimated `key' functions:}
\begin{itemize}
\item Estimate $M_c(t)$ for each $t=S,S+1,\dots, w;\ c=c_1,c_2,\dots,c_m$ by the least squares method: 
\[
\wh{M}_c(t) = \arg\min_{M\in \R} \l|q^{ID}_c(t,M|S) - q^{DATA}_c(t|S)\r|^2. 
\]
{\red Note that the error on the right-hand side can be zero. In this stage, our focus is on determining the value of $M$ at the specific time point $t$ where $q_c^{ID}(t;M|S) \approx q_c^{DATA}(t|S)$. In essence, we are seeking discrete observations of the latent function $M$ at the moment when the identified quantile function closely approximates the data quantile function.}

\item For each cohort $c=c_1,c_2,\ldots,c_m$, we treat the values $\{M_c(t)\,|\, t=S,S+1,\dots, w\}$ as discrete data sampled from the trajectory of $(M_c(t))_{S\le t\le w}$, and proceed to interpolate these values using continuous functions $h_l:\R+ \to \R$, which form an orthogonal basis in the space $L^2([S,w])$. In practical terms, we truncate the infinite series, limiting it to a {\red suitably} ``large" value of $L$: 
for $c=c_1,c_2,\dots,c_m$ and ${\red \a_c = (\a_{c,1},\dots,\a_{c,L})}$, 
\begin{align*}
M_c(t;{\red \a_c}) := \sum_{l=1}^L {\red \a_{c,l}} h_l(t),\quad S\le t\le w. 
\end{align*}

\begin{remark}
Regarding the orthonormal basis $\{h_l\}_{l=1,2,\dots}$, the choice of the Fourier series expansion is typically suitable for handling periodic data, whereas $B$-splines are preferred for non-periodic data. In our research, we use the $B$-spline expansion method.
\end{remark}

The coefficient ${\red \a_{c,l}}$ in each summand is estimated by the  least squares method:  
\[
{\red \wh{\a}_c} := ({\red \wh{\a}_{c,1},\dots,\wh{\a}_{c,L}})=\arg\min_{\a \in \R^L} \sum_{t=S}^{w} \l| M_c(t;\a) - \wh{M}_c(t)\r|^2. 
\]
\end{itemize}

In this way, we have $m$-discrete obervations as functional data for the `key' function $M_c$ in ID-SEM, and 
we obtain a basis expnsion of {\it centerd data} $\wh{M}^*_c = (\wh{M}^*_c(t))_{S\le t\le w}$ such that 
\begin{align}
\wh{M}^*_c(t) = M_c(t;{\red \wh{\a}_c}) - \ol{M}(t) =: \sum_{l=1}^L {\red \wh{\a}^*_{c,l}} h_l(t),  \label{fdM}
\end{align}
where $\ol{M}(t) = \frac{1}{m}\sum_{c=c_1}^{c_m} M_c(t;{\red \wh{\a}_c}) $, and ${\red \wh{\a}^*_{c,l} = \wh{\a}_{c,l} - \frac{1}{m}\sum_{c=c_1}^{c_m} \wh{\a}_{c,l}}$. 

Similarly, we also have centred functional data of $\wh{\Lambda}^*_c = (\wh{\Lambda}^*_c(t))_{S\le t\le w}$ in IG-SEM, where 
\begin{align}
\wh{\Lambda}^*_c(t) =  \Lambda_c(t;{\red \wh{\b}_c}) -  \ol{\Lambda}(t) := \sum_{l=1}^L {\red \wh{\b}^*_{c,l}} h_l(t),   \label{fdL}
\end{align}
with ${\red \wh{\b}_c = (\wh{\b}_{c,1},\dots,\wh{\b}_{c,L})}$ and ${\red \wh{\b}^*_{c,l} = \wh{\b}_{c,l} - \frac{1}{m}\sum_{c=c_1}^{c_m} \wh{\b}_{c,l}}$. 

\subsection{Functional principle component analysis}\label{sec:fpca}

{\it Functional Principal Component Analysis (FPCA)} is an extension of classical Principal Component Analysis (PCA) to handle random elements that take on functional values.

Consider a stochastic process $Y_c=(Y_c(t))_{t\in [0,T]}$ with continuous paths, where $T>0$. These processes are indexed by $c=1,2,\dots,m$, and their values reside in a Hilbert space $L^2([0,T])$. We assume that the random sequence $Y_1, Y_2,\dots, Y_m$ comes from weakly stationary function-valued time series, with a mean function $\mu(t) = \E[Y_1(t)]$. In this context, the covariance kernel, denoted as $K(t,s):= \Cov(Y^c(t),Y^c(s))$, is invariant with respect to $c$. It has a spectral-type decomposition, which can be characterized using {\it Mercer's Theorem} (see, e.g., Riesz and Sz.-Nagy \cite{rs90}): 
\[
K(t,s) = \sum_{j=1}^\infty \la_j e_j(t)e_j(s),\quad \la_1\ge \la_2\ge \dots \ge \la_n\ge \dots, 
\]
where $\la_j$ is the $j$th eigenvalue and $e_j$ is the corresponding {\red eigenfunction} in $L^2([0,T])$ space.
Then the {\it Karhunen-L\`oeve expansion} allows us to decompose the random function $Y_c(\cdot)$ as 
\begin{align*}
Y^*_c(t) = \sum_{j=1}^\infty {\red Z_{cj}}e_j(t),\quad {\red Z_{cj}}= \langle Y_c-\mu, e_j\rangle_{L^2([0,T])}, 
\end{align*}
where $Y^*_c(t) := Y_c(t) - \mu(t)$ and ${\red Z_{cj}}$ is the inner product in $L^2([0,T])$ of $Y^*_c$ and $e_j$: 
\[
{\red Z_{cj}}:= \int_0^T Y^*_c(t) e_j(t)\,\df t, 
\]
which corresponds to the {\it principal component score} in the classical (finite-dimensional) PCA. 
According to this manner, we can assume that 
\begin{align}
Y_c(t) = \mu(t) + \sum_{j=1}^L {\red Z_{cj}}e_j(t) + \e_c(t), \label{KL-exp}
\end{align}
where $L$ is an integer suitably chosen and $\e_c$ is a noise process depending on the index $c$. 

Our aim is to find an approximation of the expansion \eqref{KL-exp}. First, we shall estimate the covariance kernel $K$ by a sample version: 
\[
\wh{K}(s,t) = \frac{1}{m-1}\sum_{c=1}^m Y^*_c(s)  Y^*_c(t) , 
\]
and we would like to find the eigenvalues $\la_j$'s and eigenfunctions $e_j$'s of {\red $K$}: 
\[
\int_0^T K(s,t) e_j(s)\,\df s = \la_j e_j(t),\quad t\in [0,T]. 
\] 
This practical procedure is seen in Ramsay and Silverman \cite{rs05}, Section 8.4.2, or a more general argument is found in Hyndman and Ullah \cite{hu06}.  
With the expansion of our data using an orthonormal system, as detailed in \eqref{fdM} and \eqref{fdL}, we can now consider our data in a centered form, expressed as follows: 
\[
\wh{Y}^*_c(\cdot) := Y_c(\cdot) - \ol{Y}(\cdot) =\sum_{l=1}^L {\red \wh{\a}^*_{c,l}}\,\xi_j(\cdot), \quad c=1,2,\dots,m, 
\]
where $\ol{Y}(t) =\frac{1}{m}\sum_{c=1}^m Y_c(t) $, and $\{\xi_j\}_{j=1\dots}$ is an orthonormal basis in $L^2([0,T])$. 

Let $\mb{A} = ({\red \wh{\a}^*_{c,l}})_{1\le c\le m;1\le l\le L}$ is the $m\times L$ matrix of coefficients. 
Following the manner of Ramsay and Silverman \cite{rs05}, the eigenfunctions $\wh{e}_j$'s of the kernel function $\wh{K}$ has the following basis expansion: 
\[
\wh{e}_j(t) = \sum_{k=1}^L b^j_k \xi_k(t),  \quad j=1,2,\dots,L, 
\]
where $\mb{b}^{(j)}=(b^j_1,\dots,b^j_L)$ is the $j$th eigenvector of $(m-1)^{-1}\mb{A}^\top \mb{A}$. 
The corresponding eigenvalue becomes $\wh{\la}_j$: 
\[
\frac{1}{m-1}\mb{A}^\top \mb{A} \mb{b}^{(j)} = \wh{\la}_j \mb{b}^{(j)}. 
\]
Then, the estimated principal component score is given by 
\begin{align*}
{\red \wh{Z}_{cj}} = \int_0^T \wh{Y}^*_c(t) \wh{e}_j(t)\,\df t,  \quad c=1,2,\dots,m. 
\end{align*}
As a consequence, we have an approximation of \eqref{KL-exp} as follows: for $L\in \N$ large enough, 
\begin{align*}
\wh{Y}^*_c(t) =  \sum_{j=1}^L {\red \wh{Z}_{cj}}\,\wh{e}_j(t). 
\end{align*}

For the theoretical details for FPCA and the inference, see also Horv\'ath and Kokoszka \cite{hk12}, or Hsing and Eubank \cite{he15}. 

\subsection{Prediction of the key function for future cohorts}\label{sec:predict}
We implement the procedure outlined in the preceding section, applying it to the functional data $Y_c=M_c$ and $Y_c=\Lambda_c$. 
These datasets take the form:
\[
\{\wh{M}^*_{c_1}, \dots, \wh{M}^*_{c_m}\}\quad \mbox{and}\quad \{\wh{\Lambda}^*_{c_1}, \dots, \wh{\Lambda}^*_{c_m}\}
\]
as in \eqref{fdM} and \eqref{fdL}, respectively. 
We explain mainly the procedures of estimation and prediction for $M_c$.  
It is important to note that the procedures for $\Lambda_c$ follows precisely the same way. 

Let us start by recalling that we have the smoothly estimated  functional data: 
\[
\{\wh{M}^*_c(t)\,|\, t= S, S+1,\dots w;\ c = c_1,c_2,\dots,c_m\}.
\]

\noindent {\bf Step 1: Approximating the Karhunen-L\`oeve expansion.} \vspace{1mm}

Following the steps in the previous subsection, we have the eigenfunctions $\wh{e}_{1j}$'s and the corresponding eigenvalues $\la_{1j}$'s of 
\[
\wh{K}_1(s,t) =  \frac{1}{m-1}\sum_{c=c_1}^{c_m} \wh{M}^*_c(s)  \wh{M}^*_c(t). 
\]
Without loss of generality, we may assume that 
$\wh{\la}_1\ge \wh{\la}_2\ge \dots \ge \wh{\la}_L$. Then, we have 
\[
\wh{M}^*_c(t) = \sum_{j=1}^L {\red \wh{Z}^{(1)}_{cj}} \wh{e}_{1j}(t),
\]
where ${\red \wh{Z}^{(1)}_{cj}}= \int_0^T \wh{M}^*_c(t)\wh{e}_{1j}(t)\,\df t$ is the $j$th FPC-score. \
The number $L$ is numerically chosen as $L= \mathrm{rank} \wh{K}_1$, where $\wh{K}_1 = (\wh{K}_1(t,s))_{t,s=S,S+1,\dots,w}$, 
and we can reduce it so that the {\it contribution rate} of the score exceeds a certain threshold $\th\in (0,1)$ (usually close to 1): 
\begin{align}
K_{1\th}= \inf\l\{k \in \N\,\Big|\, \frac{\sum_{j=1}^k \wh{\la}_{1j}}{ \sum_{j=1}^{L} \wh{\la}_{1j} } \ge \th\r\}.  \label{CRate}
\end{align}
Then the estimated (approximated) $M_c$ is given by 
\begin{align*}
\wh{M}_c(t;{\red \wh{\a}_c}) \approx  \ol{M}(t) + \sum_{j=1}^{K_{1\th}} {\red \wh{Z}^{(1)}_{cj}} \wh{e}_{1j},\quad c=c_1,\dots,c_m. 
\end{align*}

Following the identical procedure used for $\wh{M}_c(t;{\red \wh{\a}_c})$, we arrive at an approximation for $\Lambda_c$ as follows:
\begin{align*}
\wh{\Lambda}_c(t;{\red \wh{\b}_c}) \approx  \ol{\Lambda}(t) + \sum_{j=1}^{K_{2\th}} {\red \wh{Z}^{(2)}_{cj}} \wh{e}_{2j}(t),\quad c=c_1,\dots,c_m.
\end{align*}
where ${\red \wh{Z}^{(2)}_{cj}}= \int_0^T \wh{\Lambda}^*_c(t)\wh{e}_{2j}(t)\,\df t$, $\wh{e}_{2j}$'s are the eigenfunctions of 
$\wh{K}_2(s,t) =  \frac{1}{m-1}\sum_{c=c_1}^{c_m}\wh{\Lambda}^*_c(s)  \wh{\Lambda}^*_c(t). $
The selection of $K_{2\th}$ is determined in a manner similar to that outlined in \eqref{CRate}. 

\

\noindent {\bf Step 2: Prediction of mortality function.}\vspace{1mm}

We now possess a set of time series data for the principal component scores, represented as $\{{\red \wh{Z}^{(1)}_{c_1 j}, \wh{Z}^{(1)}_{c_2 j}, \dots, \wh{Z}^{(1)}_{c_m j}} \}$. Considering this dataset as historical data, our objective is to make predictions for $\{ {\red \wh{Z}^{(1)}_{cj}} \,|\, c> c_m\}$ for each $j=1,\dots, K_{1\th}$.
To accomplish this, we will employ time series models and forecast future values based on the distribution of these scores. This approach aligns with the methodology described in Hyndman and Ullah \cite{hu06}. 
\begin{remark}
Subsequently, we will employ the ARIMA model to fit the data and utilize the 95\% prediction intervals to forecast the SEM mortality function for future cohorts. The process of implementing ARIMA predictions is straightforward, particularly when using the R package {\tt forecast}. 
\end{remark}

Once we obtain the predictor $\{ {\red \wh{Z}^{(1)}_{cj}} \,|\, c> c_m\}$ for each $j=1,\dots,K_{1\th}$, 
we have the (conditional) mortality functions of future cohorts $c>c_m$: 
\begin{align*}
q_c^{ID}(t,\wh{M}_c(\cdot;{\red \wh{\a}_c})|S),\quad c=c_{m+1}, c_{m+2}, \dots.  
\end{align*}

Regarding IG-SEM, we take the same procedures as above. 
We employ ARIMA models to forecast the future values of $\{{\red \wh{Z}^{(2)}_{cj}} \,|\, c>c_m\}$ for each $j=1,\dots, K_{2\th}$ and obtain 
\begin{align*}
q_c^{IG}(t,\wh{\Lambda}_c(\cdot;{\red \wh{\b}_c})|S),\quad c=c_{m+1}, c_{m+2}, \dots.  
\end{align*} 

\ 

\noindent {\bf Step 3: Modification of predicted mortality function.}\vspace{1mm}

This modification step is a specialized technique unique to the SEM approach and deviates from conventional forecasting procedures. In this regard, it represents a novel contribution to the field.

When performing our estimations, we utilize (empirical) mortality data up to the age of $w$ for the $c_m$-cohort. Therefore, when we seek to estimate mortality for a cohort $c\,(> c_m)$, individuals born in that cohort have already reached the age of $c_m+w-c$, and we have access to mortality data up to this age. As a result, we can draw upon information from actual mortality data up to the age of $c_m+w-c$ for the cohort $c$ under consideration to enhance our predictions.

In the previous step, where we predicted $\{ {\red \wh{Z}^{(k)}_{cj}} \,|\, c> c_m\}$ using the ARIMA model, we also obtained the corresponding prediction intervals. These intervals, denoted as $I_{\d}^{c,k,j}$, represent the $\d$-prediction intervals:
\[
\P\l( {\red \wh{Z}^{(k)}_{cj}} \in I_{\d}^{c,k,j} \r) = \d,\quad (c> c_m;\ k=1,2;\ j=1, \dots, K_{k\th})
\]
For instance, with $\d=0.95$, the true score $Z^c_{kj}$ falls into the $I_{\d}^{c,k,j}$ with a probability of $\d$. 
This assumption allows us to fine-tune the fitting of the mortality function, as exemplified below for the case where $k=1$:
\[
({\red \wt{Z}^{(1)}_{c1},\dots, \wt{Z}^{(1)}_{cK_{1\th}}})  = \arg\min_{{\red \wh{Z}_c}} \sum_{t=S}^{c_{m+w-c}}\l|q_c^{ID}(t,\wh{M}_c(\cdot;{\red \wh{\a}_c})|S) - q_c^{DATA}(t|S) \r|^2, 
\]
where $\arg\min_{{\red \wh{Z}_c}}$ is taken over all ${\red \wh{Z}^{(1)}_c}$ satisfying that  
\[{\red \wh{Z}^{(1)}_c:=(\wh{Z}^{(1)}_{c1},\dots, \wh{Z}^{(1)}_{cK_{1\th}}) \in I_{\d}^{c,1,1} \times \dots \times I_{\d}^{c,1,K_{1\th}}}.  
\]
Thus, we have the modified version of the conditional mortality function: 
\begin{align}
q_c^{ID}(t,\wt{M}_c^\d|S),\quad c=c_{m+1}, c_{m+2}, \dots, c_{m+w-S}.  \label{q-ID-m}
\end{align}
with the modified `key' function
\[
\wt{M}_c^\d(t) =  \ol{M}(t) + \sum_{j=1}^{K_{1\th}} {\red \wt{Z}^{(1)}_{cj}} \wh{e}_{1j}(t). 
\]
In the same way, we also have 
\begin{align}
q_c^{IG}(t,\wt{\Lambda}_c^\d|S),\quad c=c_{m+1}, c_{m+2}, \dots, c_{m+w-S}.  \label{q-IG-m}
\end{align}
with the modified `key' function
\[
\wt{\Lambda}_c^\d(t) =  \ol{\Lambda}(t) + \sum_{j=1}^{K_{2\th}} {\red \wt{Z}^{(2)}_{cj}} \wh{e}_{2j}(t). 
\]
\begin{remark}
The above modification step is applicable to cohorts $c=c_{m+1}, c_{m+2}, \ldots, c_{m+w-S}$ because no mortality data are available beyond the age of $S$ years for cohorts beyond $c_{m+w-S}$.
\end{remark}

\section{Data analysis}\label{sec:data}

In this section, we leverage the modified mortality functions described in \eqref{q-ID-m} and \eqref{q-IG-m} to make predictions for selected future cohorts. In our comparative analysis, we evaluate not only these two modified approaches but also consider parametric models proposed in previous works by Shimizu {\it et al.} \cite{s20, s23}.

\subsection{Swedish mortality}\label{sec:sweeden}

The Swedish mortality data available in the Human Mortality Database are among the oldest records, rendering them well-suited for long-term projections. We make use of empirical mortality functions $q_c^{DATA}(t)$ for cohorts $c_1=1781, c_2=1782, \ldots, c_m=1830$ (with $m=50$) at ages $t=20,21, \ldots, 110$ (with $S=20$ and $w=110$). Our objective is to forecast future mortality functions $q_c^{ID}(t|S)$ and $q_c^{IG}(t|S)$ for cohorts $c = 1870, 1890$, and $1910$ (which correspond to 40, 60, and 80 years following the last cohort, $c_m=1830$).

It is important to note that, as discussed in Remark \ref{rem:data}, we already have mortality data available up to the year $c_m+w = 1940$ (for the purpose of this analysis, we assume the current year to be 1940). Consequently, individuals born in cohort $c$ are already $(1940-c)$ years old, and our forecasting takes into account the age $(1940-c)+1$, assuming the individual has lived up to age $S$. For example, when predicting the cohort $c=1910$, our analysis pertains to individuals over the age of 31, and we compare the curve $q_c(t|30)$ (for $31 \le t \le 110)$.

The parametric models for key functions used for comparison below are the following setup, similar to Shimizu {\it et al.} \cite{s23}:
\begin{itemize}
\item ID-SEM \eqref{ID}: $U_c(t,\th_c) =  \a_c  + \b_c \exp\l( \g_c (t- T_c) \r)\I_{\{t \ge T_c\}}$, where $\th_c=(\a_c,\b_c,\g_c)$, 
and $T_c$ is a change point where the mortality changes drastically, fixed at $T_c={\red 50}$. 
Note that it is too hard to estimate and predict $T_c$ in practice; see Shimizu {\it et al.} \cite{s20}. 

\item IG-SEM \eqref{IG}: $\Lambda_{\th_c} (t) = e^{a_c t} + b_c t - 1$, where $\th_c=(a_c,b_c,\s_c)$. 
\end{itemize}
All parameter estimation and forecasting methods were performed according to the methodologies by Shimizu {\it et al.} \cite{s23}.

Subsequently, we maintain the values of $x_c$ representing the initial energy, $\k_c$ in \eqref{kappa}, and $\s_c$ in \eqref{IG-q} at a fixed set of values as follows:
\begin{align}
x_c \equiv 1000;\quad \k_c \equiv -0.25;\quad \s_c\equiv 0.001, \label{hyper}
\end{align}
These specific values have been empirically selected based on insights obtained from data analysis, as detailed in Shimizu {\it et al.} \cite{s23, s23-2}. They are chosen to ensure that the estimation and optimization processes remain numerically stable across all cohorts.

\subsection{Forecasting of $M_c$ and $\Lambda_c$}
\begin{figure}[htbp]
\center 
\begin{tabular}{c}
\includegraphics[width=12cm]{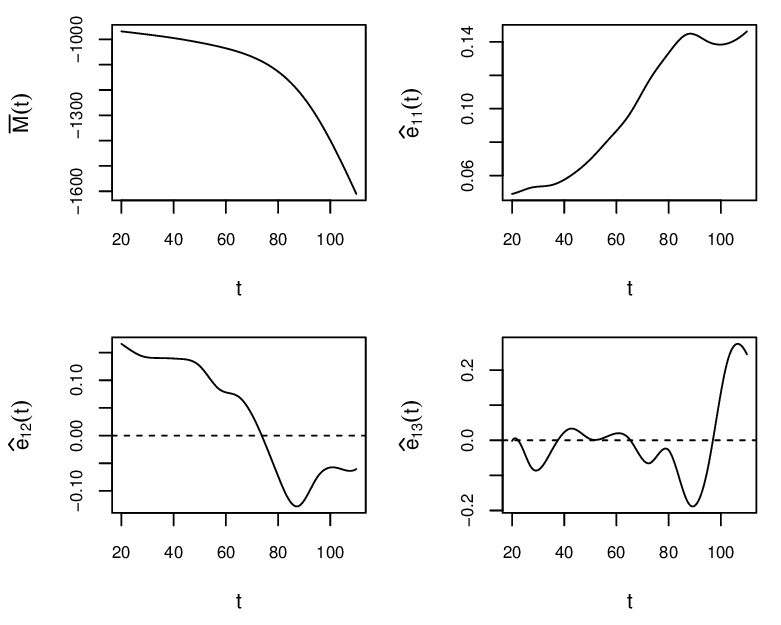} \\ (a) \vspace{3mm}\\
\includegraphics[width=12cm]{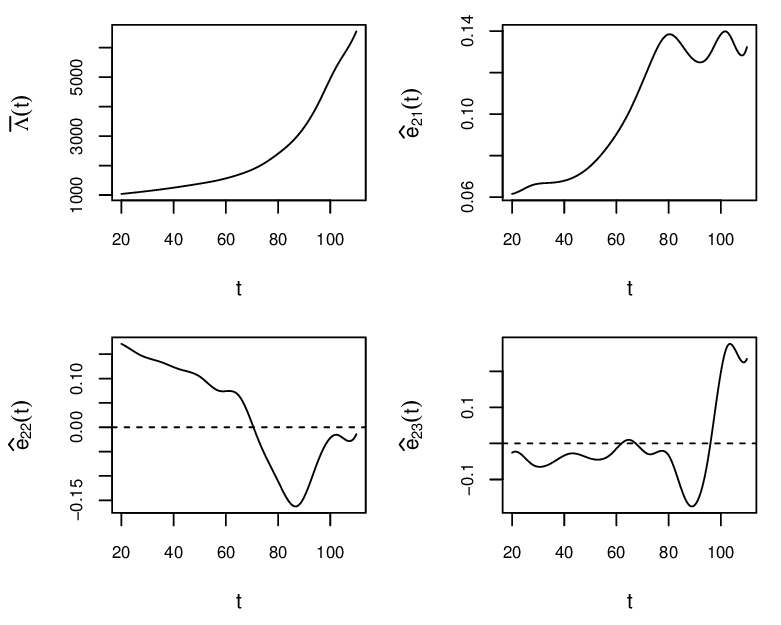} \\(b)
\end{tabular}
\caption{Figure (a) is the estimated mean functions $\ol{M}(t)$ and eigenfunctions for $M^*_c(t;{\red \wh{\a}_c})$. Figure (b) is the one for $\ol{\Lambda}_c(t)$ and $\Lambda^*_c(t;{\red \wh{\b}_c})$. }
\label{fig:pca}
\end{figure}

We employed the R package {\tt fda} to estimate the {\red eigenfunctions} $\wh{e}_{kj}$ and eigenvalues $\wh{\la}_{kj}$ $(k=1,2)$. In determining the number of terms, $K_{1\th}$ in equation \eqref{CRate}, we set $\th=0.995$, which yielded $K_{1\th} = 3$.
The output includes the mean function $\mu(t) =\ol{M}(t)$ and the eigenfunctions $\wh{e}_{1j}$ for $j=1,2,3$, as displayed in Figure \ref{fig:pca} (a). 
The same procedure for $\Lambda_c$ obtains Figure \ref{fig:pca} (b).

\begin{figure}[htbp]
\begin{tabular}{c}
\begin{minipage}{0.5\hsize}
\center \includegraphics[width=7.0cm]{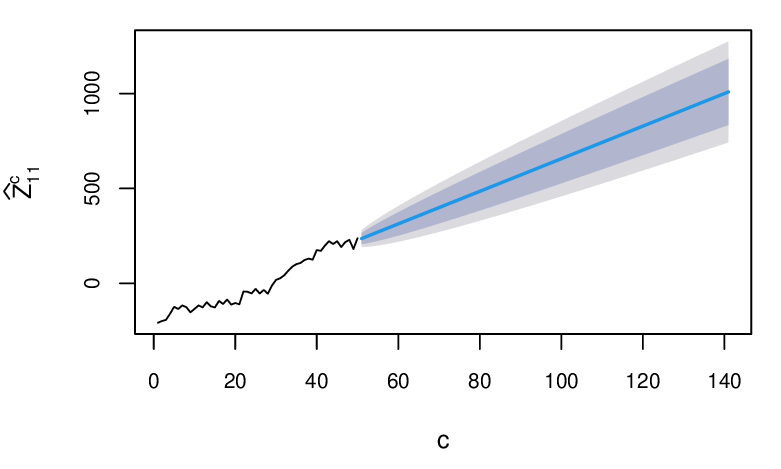} \\(a)
\end{minipage}
\begin{minipage}{0.5\hsize}
\center \includegraphics[width=7.0cm]{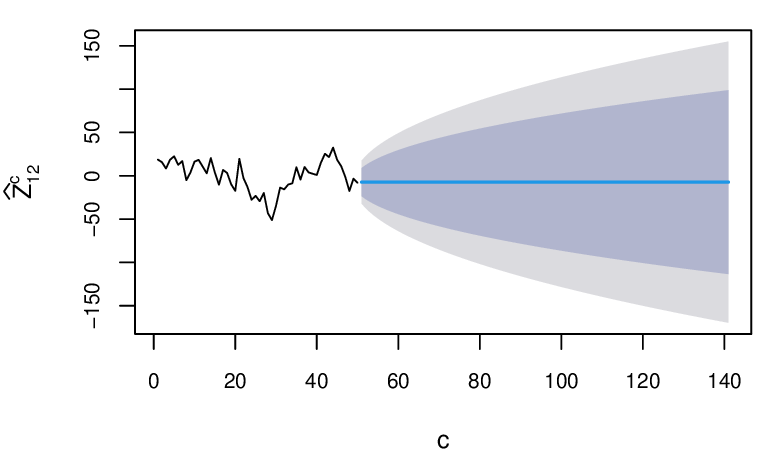} \\(b) 
\end{minipage}
\end{tabular}
\vspace{3mm} \\
\begin{tabular}{c}
\begin{minipage}{0.5\hsize}
\center  \includegraphics[width=7.0cm]{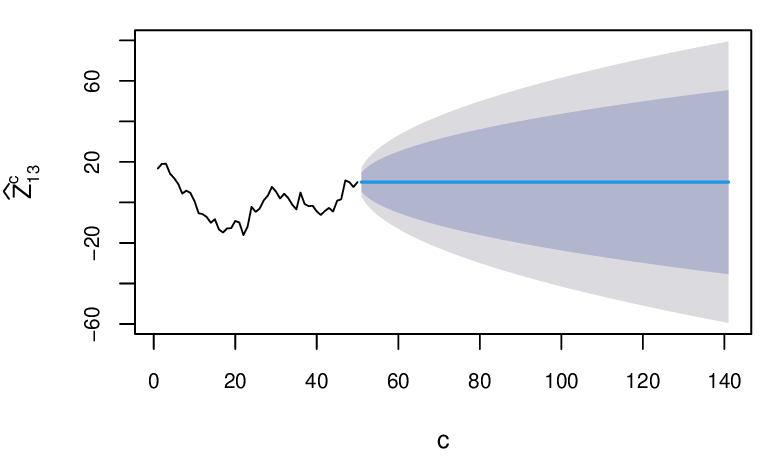} \\(c) 
\end{minipage}
\begin{minipage}{0.5\hsize}
\center \includegraphics[width=7.0cm]{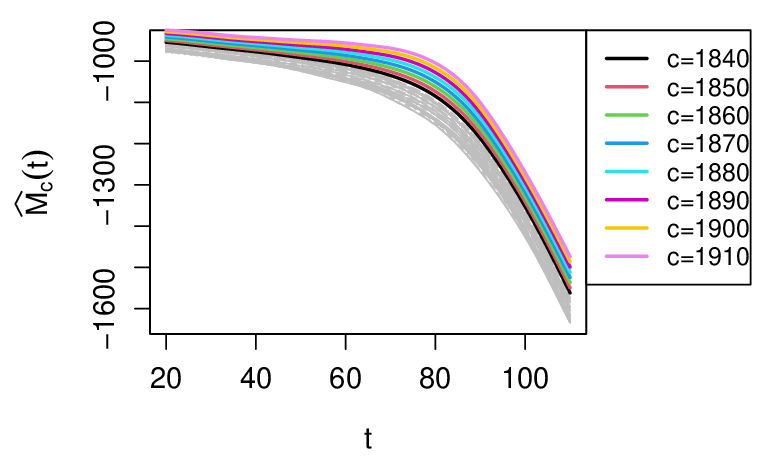} \\(d)
\end{minipage}
\end{tabular}
\caption{Figures (a)-(c) illustrate the predictions for the 1st to 3rd PCA scores. In each figure, the blue line represents the mean, while the purple and grey ranges denote the 80\% and 95\% prediction intervals, respectively. Figure (d) provides the forecast for $\wt{M}^\d_c$ for cohorts $1840 \le c \le 1910$ with $\d=0.95$. The grey curves correspond to the estimated past curves $\wh{M}_c$ for cohorts $1781 \le c \le 1830$. }
\label{fig:M-score}
\end{figure}

The predictions for the PCA scores ${\red \wh{Z}^{(1)}_{cj}}$ for cohorts $c > c_m$ are conducted using {\tt forecast} with ARIMA models. The results of these predictions are given in Figure \ref{fig:M-score}, (a)-(c), and the forecasted trajectories of $\wt{M}^\d_c(t)$ modified based on the prediction intervals are illustrated in (d).

To obtain a {\it modified} key function, $\wt{M}^\d_c(t)$, we employ the $\d=95$\%-prediction interval (represented by the grey range in the {\red Figure \ref{fig:M-score}} (a)-(c)). The graphs reveal that the function for $M_c$ exhibits a tendency to increase year by year. Given that $M_c$ originally represents a drift of survival energy, this suggests that the trajectory of survival energy is on an upward trajectory year by year. Consequently, future cohorts are less likely to experience mortality.

\begin{figure}[htbp]
\begin{tabular}{c}
\begin{minipage}{0.5\hsize}
\center \includegraphics[width=7.0cm]{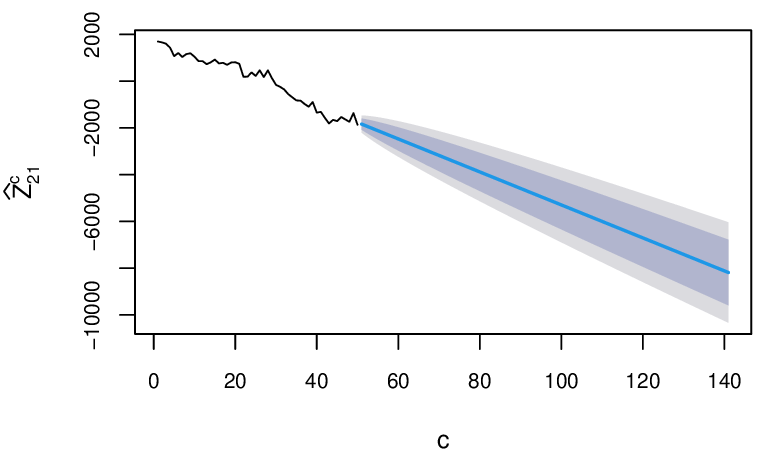} \\(a)
\end{minipage}
\begin{minipage}{0.5\hsize}
\center \includegraphics[width=7.0cm]{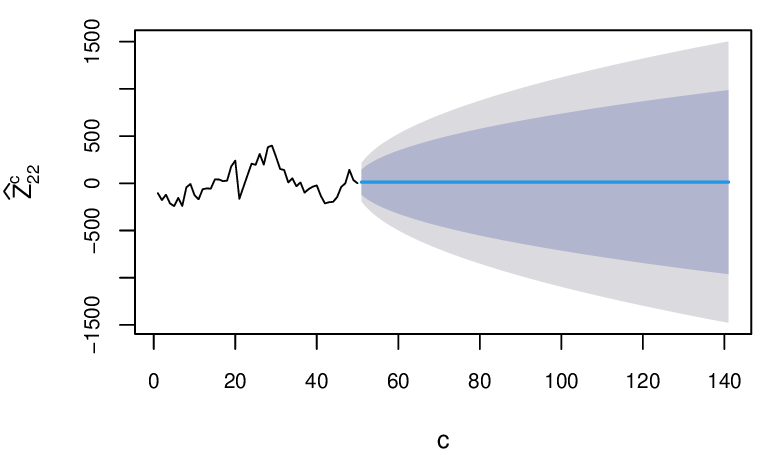} \\(b) 
\end{minipage}
\end{tabular}
\vspace{3mm} \\
\begin{tabular}{c}
\begin{minipage}{0.5\hsize}
\center  \includegraphics[width=7.0cm]{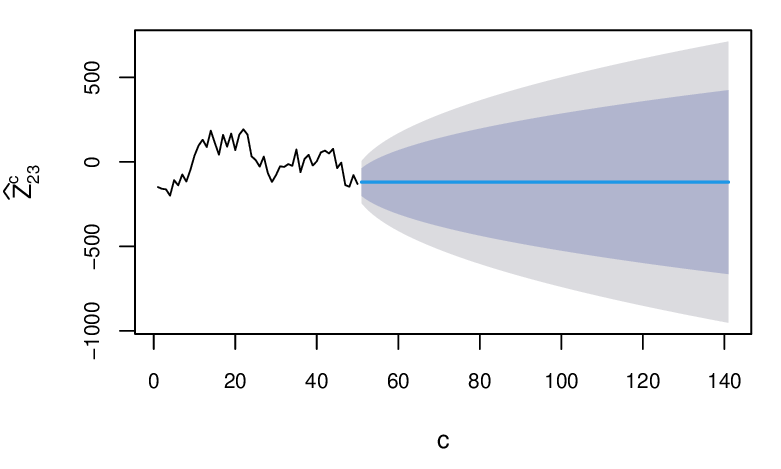} \\(c) 
\end{minipage}
\begin{minipage}{0.5\hsize}
\center \includegraphics[width=7.0cm]{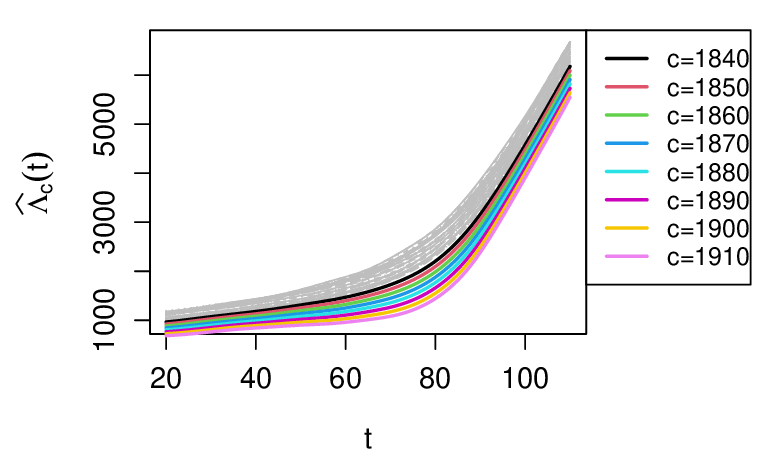} \\(d)
\end{minipage}
\end{tabular}
\caption{Figures (a)-(c) illustrate the predictions for the 1st to 3rd PCA scores. In each figure, the blue line represents the mean, while the purple and grey ranges denote the 80\% and 95\% prediction intervals, respectively. Figure (d) provides the forecast for $\wt{\Lambda}^\d_c$ for cohorts $1840 \le c \le 1910$ with $\d=0.95$. The grey curves correspond to the estimated past curves $\wh{\Lambda}_c$ for cohorts $1781 \le c \le 1830$.  }
\label{fig:L-score}
\end{figure}

Similarly, we employ analogous procedures to forecast $\Lambda_c$, and the corresponding results are presented in Figure \ref{fig:L-score}.

\subsection{Parametric vs. Nonparametric}\label{sec:vs}

\begin{figure}[htbp]
\center \includegraphics[height=6cm]{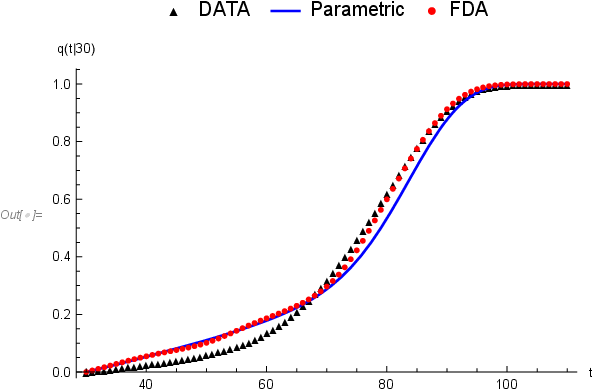} 
\caption{Comparison between the Swedish Male ID-SEM $q_{1910}^{ID}(t|30)$ predictions using parametric (Shimizu {\it et al.} \cite{s23}) and nonparametric (FDA) methods. It's evident that the FDA approach outperforms the parametric one, particularly in predicting the elderly population.}
\label{fig:ID}
\end{figure}

\begin{figure}[htbp]
\center \includegraphics[height=6cm]{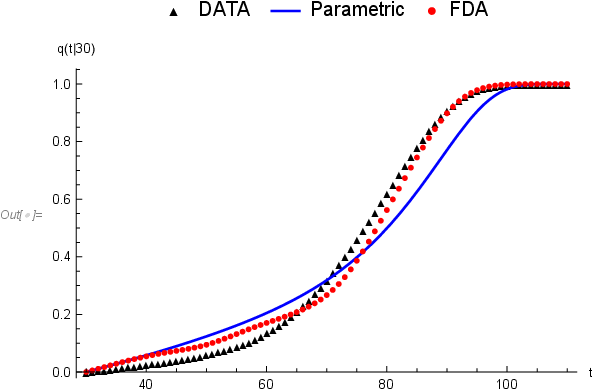} 
\caption{Comparison between the Swedish Male IG-SEM $q_{1910}^{IG}(t|30)$ predictions using parametric (Shimizu {\it et al.} \cite{s23}) and nonparametric (FDA) methods. In this case, the FDA method outperforms the parametric one overall.}
\label{fig:IG}
\end{figure}

Using the key function $M_c$ derived from the preceding step, we compute a predicted mortality function and compare it with the results of the parametric model proposed by Shimizu et al. \cite{s23} (using the same model setup as in their paper). Our predictions are cohorts born in $c=1870, 1890$, and $1910$. In our data configuration, we used data up to the cohort born in 1830 (covering ages 20-110), placing us in the year 1940. Consequently, our predictions target individuals from the $c=1870, 1890$, and $1910$ cohorts at ages 70+, 50+, and 30+, respectively.
Figures \ref{fig:ID} and \ref{fig:IG} display $q^{ID}_c(t|30)$ and $q^{IG}_c(t|30)$ for the male cohort born in $c=1910$, which represents the longest projection. Additionally, Tables \ref{tab:M-mse} and  \ref{tab:F-mse} provide a comparison of the Mean Squared Error (MSE) of males and females predicted mortality functions for these cohorts: {\red for $c=1870, 1890, 1910$ with $\wt{c}=1940-c$,  and for $w=110$, 
\[
MSE = \frac{1}{w-\wt{c}+1}\sum_{t = \wt{c}}^{w} \l|q_c^{**}(t|30) - q_c^{DATA}(t|30)\r|^2,\quad **=ID\ \  \mbox{or}\ \  IG
\]
}
The outcomes clearly indicate that our FDA approach outperforms the parametric models (except for a few rare cases) and offers superior accuracy in mortality prediction. 
As illustrated in Figures \ref{fig:ID} and \ref{fig:IG}, our methodology demonstrates a propensity for heightened accuracy in mortality prediction, particularly when applied to the elderly demographic. Furthermore, this nonparametric approach eliminates the need to select change points $T_c$ required in ID-SEM. 
This constitutes a significant practical contribution, as accurate forecasting for the elderly demographic carries substantial overarching significance.

Occasionally, ID-SEM outperforms the FDA approach in terms of predictions (e.g., 1890-cohort for females). Nevertheless, the selection of the hyperparameter $T_c$ in ID-SEM bears significant importance. An erroneous choice in this regard can lead to a sudden deterioration in predictions, rendering the parametric approach notably unstable.

\begin{remark}\label{rem:overlearn}
As discussed by Shimizu et al. \cite{s23}, the introduced "Modification" in Section \ref{sec:predict} does not invariably lead to improved predictions. This is attributed to the risk of entering an "over-learning" state, induced by excessive LSE-fitting to existing data, which subsequently diminishes the model's predictive capability for future data. One potential remedy involves the reduction of the prediction threshold, e.g., $\d =0.8$, etc. Nevertheless, based on our extensive experience in numerous data analyses, it is evident that, in most cases, this "Modification" indeed results in a reduction of MSE in the predictions.
\end{remark}

\begin{table}[h]
\center
\begin{tabular}{c|cc|cc}
\hline
Males&ID-SEM &  & IG-SEM&  \\ \hline 
Cohort & Parametric & FDA & Parametric & FDA \\ \hline
1870 & 6.6747E-4 & {\bf 9.6269E-5}&5.1827E-4 &{\bf 7.3466E-6} \\ \hline
1890 & 3.4059E-3 & {\bf 3.2618E-3}&2.6057E-2 &{\bf 4.7426E-4} \\ \hline
1910 & 1.8461E-3 & {\bf 7.6211E-4}&4.8067E-2 &{\bf 1.2011E-3} \\ \hline
\end{tabular}
\caption{Parametric vs. FDA (Males): MSEs for predictions of Swedish mortality functions of cohort $c=1870, 1890$ and $1910$, in ID-SEM and IG-SEM. }
\label{tab:M-mse}
\end{table}

\begin{table}[h]
\center
\begin{tabular}{c|cc|cc}
\hline
Females&ID-SEM &  & IG-SEM&  \\ \hline 
Cohort & Parametric & FDA & Parametric & FDA \\ \hline
1870 & 9.7386E-4 & {\bf 2.8153E-4}&1.1994E-3 &{\bf 2.1915E-4} \\ \hline
1890 & {\bf 1.0991E-4}& 1.2383E-4&1.8389E-2 &{\bf 1.1757E-3} \\ \hline
1910 & 5.8036E-3 & {\bf 1.0184E-3}&1.8928E-2 &{\bf 1.1845E-2} \\ \hline
\end{tabular}
\caption{Parametric vs. FDA (Females): MSEs for predictions of Swedish mortality functions of cohort $c=1870, 1890$ and $1910$, in ID-SEM and IG-SEM. }
\label{tab:F-mse}
\end{table}


\section{Concluding remarks}\label{sec:remark}

We used a nonparametric approach with functional data analysis for cohort-specific mortality rate prediction based on Survival Energy Modeling (SEM). In contrast to existing research with parametric models, which lack objective model selection criteria and make it difficult to identify data-driven models, our FDA approach allows for easy estimation and prediction of key functions.
In fact, this approach was able to significantly outperform predictions made by existing parametric models in both the ID-SEM and IG-SEM. 
Additionally, the convenient environment of R package for the FDA makes these approaches readily applicable in practice without the need for extensive theoretical expertise, making it a valuable tool.
While finding more powerful candidates than ID-SEM or IG-SEM for future predictions is a potential challenge, our data analysis suggests that ID-SEM already possesses sufficient predictive power at this stage.

However, amending forecasts through the utilization of the prediction interval associated with the PCA score can occasionally yield counterproductive outcomes. This issue bears a resemblance to the concept of "overlearning", where excessive adaptation to historical data results in inadequate performance when applied to future data. One potential remedy involves the reduction of the prediction threshold denoted as $\d$. Nonetheless, based on our empirical findings, even when errors worsen, they tend to be of minor magnitude, and, in the majority of instances, such adjustments lead to an enhancement in forecasting accuracy.

As discussed by Shimizu {\it et al.} \cite{s23-2}, cohort-specific mortality rate prediction with SEM can be a powerful tool for calculating insurance premiums for each cohort, predicting changes in pension amounts, and financial forecasting. Particularly in countries with increasing life expectancy, such as Japan, the analysis of longevity risk is a crucial task, and SEM is likely to become a valuable tool for addressing this challenge.

\ \vspace{5mm}\\

\noindent {\bf\large Acknowledgement.} This work is partially supported by JSPS KAKENHI Grant-in-Aid for Scientific Research (C) \#21K03358. 

\ 

\noindent {\bf\large Conflict of interest statement.} On behalf of all authors, the corresponding author states there is no conflict of interest. 

\end{document}